\documentclass[aps,showpacs,toolkits,nofootinbib,preprint]{revtex4}
\newcommand{\beq}{\begin{equation}}
\newcommand{\eeq}{\end{equation}}
\newcommand{\beqs}{\begin{eqnarray}}
\newcommand{\eeqs}{\end{eqnarray}}

\newcommand{\mX}{m_X}
\newcommand{\mU}{m_U}
\newcommand{\GeV}{~ {\rm GeV}}
\newcommand{\TeV}{~ {\rm TeV}}
\newcommand{\MeV}{~ {\rm MeV}}
\newcommand{\keV}{~ {\rm keV}}

\begin{document}

\markboth{Shmuel Nussinov}
{Some Aspects of New CDM Models and CDM Detection Methods}

\title{Some Aspects of New CDM Models and CDM Detection Methods}

\author{Shmuel Nussinov}

\address{School of Physics and Astronomy, Tel Aviv University, Tel Aviv, Israel \\
Schmid College of Science, Chapman University, Orange, CA 92866}

\begin{center}
FOREWORD
\end{center}

The following is based on a mini-review which is being prepared for
Modern Physics Letters A (World Scientific Publishing Co.).


\vspace*{.5in}

\begin{abstract}

We briefly review some recent Cold Dark Matter (CDM) models. Our main focus are charge
symmetric models of WIMPs which are not the standard SUSY LSP's (Lightest Supersymmetric
Partners). We indicate which experiments are most sensitive to certain aspects of the
models.  In particular we discuss the manifestations of the new models in neutrino telescopes
and other set-ups.  We also discuss some direct detection experiments and comment on measuring 
the direction of recoil ions---which is correlated with the direction of the incoming WIMP. 
This could yield daily variations providing along with the annual modulation signatures for CDM.

\end{abstract}

\maketitle

\section{Introduction}

The subject of the missing dark matter is clearly part of astro-particle physics.
The astrophysics addresses the evidence for dark matter from rotation curves
\cite{Begeman:1991iy}, gravitational lensings \cite{Koopmans:2002qh}, galactic and
cluster motions (notably the colliding galaxies in the ``Bullet" cluster with data in
X-ray, visible light and gravitational lensing \cite{Markevitch:2003at}), structure
formation and WMAP measurements.\cite{Komatsu:2008hk}. These strongly suggest CDM (Cold
Dark Matter) with average local density of $\sim 0.3 GeV /(cm^3)$ and total contribution
to the cosmological energy density of $\Omega_{CDM}h^2 \sim 0.1131 \pm 0.0034$.

Here we consider the complementary particle aspect addressing the nature of the CDM
particles X, its mass $\mX$ and interactions, the models in which it arises and the
its possible direct and indirect signatures in underground or space detectors.

From both theoretical and experimental points of view the models fall into two broad
categories: ``charge" asymmetric and symmetric.

In the first category the CDM reflects an asymmetry between the number densities $n(X)$ and
$n(\bar{X})$, analogous to the baryon asymmetry. After efficient $X -\bar{X}$ annihilations
only the excess, say, $n(X)-n(\bar {X})>0$ remains. Hence there can be no signal from
present day annihilations in these scenarios.

For both categories putative decays of CDM particularly those with monochromatic
photons can provide a signature as would the underground CDM induced nuclear recoils
providing that the X-Nuclear cross section is not too small.\cite{Chivukula:1992pn}
Certain aspects of the asymmetric scenarios are rather appealing. One can invoke analogs
of the well-known baryon asymmetry despite the fact that there is no consensus re the
origin of this  asymmetry.
To explain the observed $\Omega_{CDM}/\Omega_B \sim 5-6$ one needs a CDM charge asymmetry
roughly similar to that of baryons and a matching moderate mass ratio so that:
$\Delta Y(X)/\Delta Y(B) \cdot \mX/m(B) \sim 5-6$.
This seems feasible  in a scenario where the WIMPs are neutral technibaryons with
a mass ratio  $m(TB)/m(B) \sim 1000 $ \cite{Nussinov:1985xr} and in recent SUSY CDM
variants.\cite{Kaplan:2009ag}.
Conceivably the number of $\bar{B}'$s of mass $m_{B'}\sim 5-6 \GeV$ in a ``Hidden Sector"
can be forced to equal that of ordinary baryons by an overall $B+B'$ conservation.
Another possibility with some obvious astrophysical pitfalls is to have six ``mirror
images of our S.M. and standard baryons.\cite{Foot:2007nn}
In certain asymmetric models the CDM consists of large nuggets with  $N{\rm(baryon)}
\sim 10^{20}$. These nuggets formed at the time of QCD phase transition when the
temperature was $(T \sim 100 \MeV)$  contain $ \sim$ 83\% of all baryons, thereby evading
constraints from BBN (Big Bang-Nucleosynthesis).  These nuggets consist of strange quark
matter \cite{Witten:1984rs} or other stable forms of nuclear matter.\cite{Zhitnitsky:2006vt}

In a speculative variant the initial numbers of baryons and of anti-baryons are equal.
A large $\mathcal{O}(1)$ CP violating $\theta_{\rm QCD}$ at nucleosynthesis causes
anti-baryonic nuggets to form 3:2 times more than baryonic nuggets leaving an excess
$\sim$ 1/6 of the total number of baryon+anti-baryons as the observed unclustered
baryons.~\cite{Forbes:2008zz}

This then dispenses with the need for a new particle physics sector and ``explains"
baryogenesis as well!

In this review we focus on symmetric models, mainly those in which the CDM are WIMPs,
Weakly Interacting Massive $\mX~\mathcal{O}(10-1000\GeV)$  Particles, rather than axions
and/or other relatively light ($\mX < \MeV$) long-lived bosons.~\cite{Pospelov:2008jk}
``Warm" dark matter such as sterile $\mathcal{O}(\keV)$ neutrinos were discussed at some
length. These could provide Pulsar Kicks via (spatially) asymmetric emission from
supernovae and their radiative decays (still not observed via monochromatic photons from
decays of clustered DM in haloes) may induce an early re-ionization.~\cite{Kusenko:2008gh}

Elaborating on all of this and on recent experimental advances and improved bounds on
standard axions and ``axion-like" particles as well as on conjectured MeV
bosons \cite{Boehm:2003bt} will carry us too far from the limited mission of this review.

There is, however, one interesting exception. Solar axion to gamma conversion in single
crystals in underground detectors is enhanced when the momentum of the axions or photons
they covert to with known (solar) direction and magnitude---the (roughly measured) energy
of the photon satisfy a Bragg condition. This modulates the expected intensity at given
latitude and longitude as a function of time in a unique pattern; a feature that was used
to increases the sensitivity of solar axion detection by up to two orders of
magnitude.\cite{Avignone:2009ay}. A weak variant of this motif are daily modulations due
to channeling of the recoil ions in underground crystals.
If ever detected, these and possibly other measurements of the ion's direction will
complement the search for annual modulations suggested by Drukier, Freese, and
Spergel~\cite{Drukier:1986tm} and claimed to be seen by the DAMA/LIBRA collaboration.

\section{Some Twists on CDM Models}

In symmetric WIMP models the $X$ and $\bar{X}$ particles annihilate in the early
universe leaving relics with $Y(X)=n(X)/s =n(\bar{X})/s \sim n(X)/n(\gamma)$.
The cosmological relic density then is: $2 n(X)|_{\rm freeze-out}\times \mX$. ``Freeze-out"
occurs when the Hubble expansion rate $\sim T^2/m_{\rm Planck}$ equals the rate of
annihilations, namely $n(X)v(X)\sigma(X-\bar{X})$. The $v\sigma$ factor in the last
expression is usually (for S wave annihilations) constant as a function of energy or
temperature around $T \sim 0$.  Also before freeze-out the number density $n(X)$ is proportional
to the Boltzman factor $\exp{-m(X)/T}$. One generally finds \cite{Jungman:1995df}
that $T|_{f.o.} \sim \mX/30 - \mX/20$ and a residual cosmological CDM density $\Omega(CDM)$
proportional to $1/\sigma_{\rm ann}$.

In SUSY models with unbroken R-parity symmetry the LSP's of masses $\mX\sim \mathcal{O}
(100\GeV)$ are stable.
The fact that the annihilation $X-\bar{X}$ cross section expected for the LSP's, which are
roughly of weak interaction strength, can reproduce the desired  $\Omega_{\rm CDM}
=\Omega(X) \sim 0.2 $ appears to be a success of these SUSY models. Also WIMPs of
$\sim$ 100 GeV mass are kinematically ideal for direct detection via nuclear recoils in the
underground detectors using nuclei of similar masses. As collider experiments keep
tightening the lower bounds on the masses of Squarks,  Sleptons and Gluinos and
underground searches lower the upper bounds on the nuclear cross sections, the parameter
space for minimal SUSY models with acceptable $\Omega (X)$ keeps shrinking.  Also, Ref.
\cite{Feng:2008ya} suggested that generic cancelation between the mass of the CDM and its
couplings can generate the correct $\Omega_{\rm CDM}$ without the MSSM LSP.

In common SUSY scenarios the LSP's are ``Neutralinos": superpositions of Binos, Winos and
Higgsinos and in some classes of models the LSP's are Sneutrinos or Gravitinos.
The LSP's may be detected at the end of the decay chains of the pair of SUSY partners
hopefully to be produced at the LHC manifesting as missing transverse energy/momentum.
R parity violations or production of NLSP's, slightly more massive than and decaying to the
LSP's yield a more dramatic signature of ``Displaced" vertices correlated with the main
event with missing momentum at both the original and the displaced vertex. Longer-lived
Gluino LSP's also arise in Split SUSY scenarios \cite{ArkaniHamed:2004fb}, where the
sfermions are much heavier than the gauginos.  Also a gravitino LSP can be
long lived.\cite{Ji:2008cq}.

The WIMPs can have locally enhanced densities  (e.g., at the galactic center and much more
at the solar and earth's cores). The enhanced $X-\bar{ X} $  annihilations into positrons
and photons (or into energetic neutrinos which can escape the sun/earth) afford indirect
WIMP detection methods.  These issues have been elaborated in theses and reviews and ``Dark
SUSY" codes where aspects of SUSY models can be numerically studied \cite{Gondolo:2004sc}.

We will not study these models and focus on the more recent crop of CDM models. Various key
ingredients of the new models appeared some time ago.  This includes the ``inelastic"
\cite{TuckerSmith:2001hy}, and ``exciting" \cite{Finkbeiner:2007kk} DM models, Sommerfeld
enhancement by ($1/v$) of $v \cdot\sigma(X-\bar{X})_{\rm ann}$  due to exchange of ``light"
particles \cite{Hisano:2004ds,Cirelli:2005uq,ArkaniHamed:2008qn} and the positrons from the
decays of the latter~\cite{Cholis:2008vb}. Recently putative positron anomalies and the
possibility ``Hidden valley"  \cite{Strassler:2006im} physics be detected at the LHC instead
of or in addition to standard SUSY, launched a ``Unified" CDM model incorporating all of
these.\cite{ArkaniHamed:2008qn}  Further models where the U(1) gauge boson U is enlarged to a
non-abelian Higgs sector were studied in Refs. \cite{Baumgart:2009tn,Cheung:2009qd,Katz:2009qq}).

Finding how specific predictions of any given model manifest in specific WIMP indications in
each experiments is lengthy and redundant. Rather we address features abstracted from
(classes of) models and their manifestation in the various (classes of) experiments.

To illustrate these issues, consider the ``minimal" extended model with DM charged under
only a new abelian gauge-group. It has the basic scales of the WIMPs' mass
$\mX \sim 100 \GeV-\TeV$, the vector-boson mass at $\mU \sim \mathcal{O}(\GeV)$ and
$\Delta \mX =\mX'-\mX\sim 100 \keV$ splittings between the lowest WIMPs. Each of these
scales is phenomenologically constrained.  Thus $\Delta\mX$ was chosen to explain the
large annual modulations seen by DAMA without conflicting with bounds on WIMP---nuclear
cross sections implied by other experiments.
Also, $\mU > \MeV$ is essential to allow its $e^+ e^-$ decays to explain
PAMELA \cite{Adriani:2008zr} and ATIC \cite{:2008zzr} excesses (or the weaker yet more
solid putative anomaly)
in the positron spectrum recently found at FERMI LAT \cite{Abdo:2009zk}.

These disparate masses can be all explained via a simple pattern of radiative corrections.
Finally even very heavy fields carrying ordinary E.M. and the new U(1) charge induce a small
kinetic mixing: $\epsilon F_{\rm em}^{\mu\nu} F^\prime_{\mu\nu}$ between the new ``dark
photon" of field strength $F'$ and the ordinary photon (and also $Z$) of strength
$\epsilon \sim ee'/8\pi^2 \sim 10^{-3}$.
The $U(1)$ meson is a multi-tasking ``workhorse" achieving all the following:

\medskip

\noindent(i) Sommerfeld Enhancement (S.E.):

The $U$ exchange between $X$ and $\bar{X}$, enhances the annihilation cross section by:
$ \pi \cdot \alpha'/v$. This can be $\mathcal{O}(10 -100)$ for the small relative velocities,
say, $v<10^{-3}$ in haloes but not in the early universe with freeze-out $v \sim 1/3$).
The S.E. manifests classically in the $(v_{\rm escape}/v)^2$ enhancement of gravitational
accretion of slow particles onto a compact star.  Quantum mechanically, in the absence of
S.E. $v \sigma$---rather than $\sigma$---is constant in the low energy region. Hence the
enhancement is by $\sim 1/v$ only.\footnote{The very low velocities of relic protons and
antiprotons at and after recombination and ensuing dramatic S.E. enhancement explain the
present complete absence of slow anti-protons. (Note that the massive $U$ exchange falls
exponentially for distances greater than the Compton wavelength of the $U$ boson, and unlike
for photons, the enhanced cross section saturates at $\sigma \sim \pi \cdot (1/\mU)^2$).}

\medskip

\noindent(ii) Preferred Annihilation Channels:

Unless we postulate appreciable couplings to ordinary weak bosons $X -\bar{X}$ annihilate
predominantly into a pair of $U$'s. Via its mixing with the photon the $U$ quickly decays
to all kinematically allowed $l^+l^-$ and $q-\bar{q} $ pairs.  The latter manifest as
$\pi^+ \pi^-$ pairs, $\pi^+ \pi^-\pi^0$ triplets and as kaons and nucleon anti-nucleon pairs.
 The absence of antiproton excess at high energies \cite{Adriani:2008zq} suggests that
$\mU < 2 m_N $. Bounds on  high energy photons from the direction of the galactic center
suggests taking $\mU \sim 0.7 \GeV $ so as to avoid strong $\pi^0 $ production (followed
by $\pi^0 \rightarrow 2 \gamma$, yet allow $U$ decays to $e^+ e^- $ and $\mu ^+ \mu^-$.
~\cite{Meade:2009rb}

\medskip

\noindent(iii) WIMP-Nucleus Couplings:

The $U$-photon mixing generates coherent spin independent WIMP-nucleus interactions and
nuclear cross sections comparable to those due to $Z^0$ exchange as the $\epsilon $ factor
is compensated by $(m_Z/\mU)^2$.

If $X$ and $X'$ are Majorana components then only off-diagonal $X'XU$ couplings are
allowed. Since $X'$ and $X$ are almost degenerate this does not affect the S.E.  The $U$
decay $X'\rightarrow X + \nu + \bar{\nu}$ mediated via $U-Z^0$ mixings has a lifetime
longer than the age of the universe.~\cite{Finkbeiner:2009mi}. However, one can naturally
add to the dark sector other light bosons $b^0, \; b^{0\prime}$, etc., and the decay
of $X': \;\;X'\rightarrow X + b^0$ followed by $b^0 \rightarrow 2 \gamma$ can be made fast
enough so as to meet all observational and astrophysical constraints.

\medskip

How do all these novel aspects affect the possible experimental signatures of dark matter?
Particularly interesting is the signal of upward going muons in neutrino telescopes,
pointing either to the center of the earth or to the core of the sun. Such muons originate
from interactions in earth of the energetic neutrinos from WIMPs annihilating at the
respective cores. Unlike the case of putative electromagnetic signals from WIMP
annihilations in space, this peculiar upward-going muon signal cannot be ``faked" by any
known astronomical mechanisms. Most stages involved in generating the flux of energetic
neutrinos from the sun (or earth) are now different from the well-studied case of Neutralino
LSP's.~\cite{Jungman:1995df}

The S.E.~of $X-\bar{ X}$ annihilations enhances the signal particularly if high WIMP
densities ensuring a steady state (where every accreted WIMP eventually annihilates) are
not achieved. This is the case for WIMP accumulations in earth where the important helpful
role of S.E.~in generating the signal has been recently emphasized.~\cite{Delaunay:2008pc}

However, if the particle responsible for the Sommerfeld enhanced $X \bar{X}$ annihilations
is the above $U$, and further U is the only ``light" boson appreciably coupled to the WIMPS,
then the U-pair annihilation channel dominates and the energetic neutrino signal disappears.
The $U$'s decay into muon or pion pairs which in turn decay to neutrinos. Such energetic
neutrinos are indeed expected along with the electrons and positrons from annihilation
of halo WIMPs. However, in annihilations in the solar/terrestrial cores, the pions and muons
produced therein encounter large ($150/20 gr/{cm^3}$) densities and lose all their energies
before decaying to $\mathcal{O}(30 \MeV)$ neutrinos. Only prompt neutrinos from decays of
$Z,W, b, c$ quarks and $\tau$'s are relevant for neutrino telescopes.
A $U$ of sub GeV mass cannot decay into a pair of $\tau$'s.
Still the ``minimal" initial model can be extended to have substantial annihilations to
$Z$'s and $W$'s thereby supplying the energetic neutrinos analyzed in the JGK review.

Another kinematical issue can arise. One may wonder if the requirement of exciting in the
first collision of $X$ in the solar core the $X'$ with  $\Delta \mX= \mX'-\mX \sim 100
\keV$ decrease the fraction of such collisions which lead to gravitationally bound WIMP.
Further this $\Delta \mX$ can prevent the slow $X$ WIMPs from re-scattering into $X'$
WIMPs even on heavy, say, Iron targets, and thus impede the last stages of further WIMP
concentration.

This was studied in detail in Ref. \cite{Nussinov:2009ft} finding sufficient WIMP
concentration in the stellar core. Thus for $X-\bar{X}$ annihilations with appreciable
branchings into prompt neutrinos one can exclude using Super-Kamiokande data
\cite{Desai:2004pq}, DAMA type WIMP nuclear cross sections.

\section{DAMA/LIBRA, the Annual Modulations and Prospects of Daily Modulations Related
to Changing Recoil Directions}

Many experiments searched for nuclear recoil due to collisions of WIMPs in a wide variety
of low background underground detectors and did not find it.  The task of ``proving" that
CDM exists is daunting.\cite{Lewin:1995rx}  One
looks for {\it isolated} nuclear recoils where a WIMP of mass $\mX$ transferred to a
target  nucleus of mass $m(A,Z) \sim A$ GeV a momentum $q$. $q \sim \mu \cdot v$ where
the reduced mass $\mu$ is for a large or small ratio of masses $r= \mX/m(A,Z)$, the smaller
among the WIMP and nuclear masses, and $v \sim 10^{-3}$ is a ``virial" velocity of the
in-falling WIMP. When both masses are $\sim$ 100 GeV the transfer can be as large as 100
MeV and the corresponding recoil energy $q^2/2m(A,Z)\sim 50 \keV$.
In general, only some portion of the recoil energy, the so-called ``quenching factor"
$\sim $ 10\%   manifests in the measured electromagnetic signal. The latter signal is the
electron hole current in some Germanium semiconductor detectors or the scintillation
photons in the sodium-Iodine (Na-I) single crystals of DAMA and LIBRA. Recalling that most
H.E.~detectors have $\mathcal{O} (100 \MeV)$ thresholds helps appreciate how difficult it
is to reliably measure the resulting $\sim 5 \keV$ signals.
Also even deep underground nuclear spallation by penetrating $\mathcal{O} (\TeV) $ muons
can yield neutrons which drift towards the detectors and leave some isolated imprint there.

Recent advances in cryogenics allow also calorimetric measurements.\cite{Ahmed:2009rh}

Cooling a Germanium crystal of $\sim 100 {\rm cm}^3$ volume to very low temperatures
reduces the specific heat ($c_v \sim T^3$) so that even the tiny recoil energy deposited
measurably heats it up. By jointly measuring both the ionization energy and the total
calorimetric deposition the CDMS experiment reduces the purely E.M.~background for which
the two are equal.

Several underground experiments are carried out or planned with a different primary goal of
finding neutrinoless double beta decays in samples enriched with the proper isotopes.
However, also the IMB and Kamiokande which discovered the neutrino pulses from supernova
1987a and atmospheric neutrino oscillations, were designed to look for nucleon decay.

It is much more difficult to estimate the nuclear cross section than the $X-\bar{X}$
annihilation rates unless we commit to specific models (such as neutralino LSP's). If we
treat the nucleus as a point particle, then $d\sigma/dt  \sim G*_F^2$ with $G*_F$  an
equivalent Fermi constant (which for the spin-independent case is $\sim A$), then
$\sigma=\int_0^{t_{\rm max}} d\sigma/dt $ yields when $m({\rm WIMP}) \sim m(A,Z)$. This then
allows probing X-nucleon cross sections down to $\sigma(X-N)\sim 10^{-43}\; cm^2$.
The above is somewhat modified by nuclear form factor damping which becomes important
for $q \cdot R(A,Z) \sim 1/2 \; R(Z,A)/{\rm Fermi} \gg 1$).

Lacking a clear {\it positive!} signature for recoils due to CDM, it is extremely difficult
to claim that some excess of events seen in underground experiments above expected
backgrounds are indeed nuclear recoils due to WIMP interactions.
Some time ago the ingenious suggestion of annual modulations of the WIMP signal had been
made in Ref.~\cite{Drukier:1986tm}. The basic underlying kinematics is rather simple:
Let us make the reasonable assumption that the WIMPs and their velocities are isotropically
distributed in the rest frame of our galactic halo. The motion of the sun due to the
galactic rotation and another small radial component generate in the rest frame of the sun
a WIMP drift or ``WIMP Wind" at a fixed direction ($\sim 42^o$ with respect to the earth's
rotation axis) and velocity $v_D \sim 230 {\rm km/sec}$, comparable to the estimated virial
isotropic velocity in the halo: $<v^2>^{1/2} \sim 220 \sim {\rm km/sec}$. It turns out that
the $\sim 30 {\rm km/sec}$ velocity of the earth around the sun (more precisely, half this
value due to the $60^o$ angle between the ecliptic and galactic planes) adds to or
subtracts from $v_D$ at the beginning of June and September, respectively. Hence, we expect
that the WIMP flux and also the WIMP's energies will be maximal/minimal at these times
with a modulation amplitude of order 6\%.

Only the DAMA experiment which has been running for roughly 10 years, recently as DAMA/LIBRA
with more than double the number of the ultra-pure large Na-I crystals, claims to see this
modulation with the correct phase. On face value the rather large $\sigma (X-N)$ that the
amplitude of modulations suggests conflicts with rather stringent bounds implied by the
lack of CDM discovery in other Germanium and Xenon detectors. There have been, however,
several attempts of using modified WIMP models and the different target nuclei/materials
in order to reconcile DAMA's signal with the other experiments. Specifically:
Ref.~\cite{Bottino:2007qg} suggested relatively light WIMPs of mass $\mX < 10 \GeV$ so
that only tiny recoil energies are expected in the experiments with the heavier targets.
Yet non-negligible recoils in scattering on the relatively light sodium could be detected
by the the low threshold DAMA experiments. Note that for such light WIMPs the $X$-Na cross
section is enhanced only by $\sim A^2$ and with $A \sim 30$, the required $X-N$ cross
sections are rather high.

Smith and Weiner \cite{TuckerSmith:2001hy} suggested that the inelastic scenario can
explain DAMA if scattering with nuclei is dominated by inelastic transition into an excited
state, roughly $\mathcal{O}(100\keV)$ above the DM state. The very severe limit on
$\sigma(X-N)$ from the CDMS and Xenon and CREST experiments stem from assuming elastic
WIMP nuclear scattering and the lack of the much larger fraction of low energy events
expected in models with elastic scattering. In contrast the endothermic process in the
inelastic scenarios tends to generate larger recoils and such higher energy events are
indeed seen at some level in these experiments. Also in this scenario the higher end of
the WIMP velocity distribution which is more sensitive to the annual modulations is used.
For further discussions of all these issues see Ref.~\cite{Chang:2008gd}.

In principle, solar axions to which the other experiments are largely insensitive could also
manifest in DAMA. However, since the sun is closest during the (northern)  winter, the
modulations of such a signal are completely out of phase with the DAMA observation. In this
connection it is worthwhile to note the established $\sim 5\%$ enhancement of the TeV muon
flux underground \cite{Ambrosio:1997tc} at Grand SASSO in the summer. This increase is due
to the ``swelling up" of the hotter atmosphere so that the first hadronic interactions of
a high energy cosmic ray producing energetic pions occurs higher up in the atmosphere and
these pions can travel a longer distance in the more dilute atmosphere and have a better
chance to decay into muons before interacting.
Since the heating up effect is maximal in August rather than in June this
cannot explain DAMA. Still s repeat of DAMA in the southern hemisphere
seems vital in order to clearing lingering doubts.

One of the arguments used by the DAMA collaboration to explain why they see a signal which
others failed to see utilizes channeling of the recoil ions. Channeling operates in
crystals but not in amorphous detectors like liquid Xenon. If the recoiling ions make small
angles with respect to some Bragg planes of low Miller indices, then rather than going on
and colliding with many nuclei, the ions keep reflecting from these planes and stop after
a much longer distance.
The channeled ions tend to lose a much higher fraction of their energy via electromagnetic
processes, scintillation in DAMA's case, than unchanneled ions.
Summings over the 100, 110 and 111 planes we find that on average $\sim 25\%$ of the Iodine
or Sodium ions with low $\mathcal{O} (5 \keV)$ energy are
channeled.\cite{Drobyshevski:2007zj}

Most ($\sim 97\% $) of DAMA's  data are in the $3-6 \keV$ range. If we identify these as
the unchanneled sample then the true energies are $\sim 30-60 \keV$ and channeling effects
are much smaller. Still channeling should have generated a high energy ``echo" of the
DAMA signal. The more interesting alternative is that the signal is mainly due to the
channeled events which could be the case for low WIMP masses. In this case the daily
changing efficiency of channeling can conceivably be detected as discussed
in Ref.~\cite{Avignone:2008cw} and also briefly below.

Modulations of the ion recoil directions could be very important CDM signatures. Since
the modulated signal is due to the constant direction of the WIMP wind the variations are
according to the {\it sidereal} rather than the solar days. This makes any putative
effect robust to any systematic variations of temperature, noise, grid voltage, etc.
While reminiscent of earlier work on variation of the rate of $\gamma\rightarrow
{\rm axion}$ conversion in single crystals with the angle between the direction to the
sun and the various principle Bragg planes, the effect here is diluted in several stages:

(1) Unlike the unidirectional solar axions (all of which originate from the sun) the virial
random velocity component of the WIMPs, comparable to the drift velocity, smooths the
directional WIMP distribution making it far less peaked in the ``Wind" direction.

(2) The axion converted in static fields into a photon with exactly the same energy and
direction as the original axion. In the present case, after colliding with a WIMP the
recoil ion moves in the forward hemisphere. Yet it does not exactly follow the direction
of the initial WIMP even for the case of point-like particles or isotropic (pure S wave)
collisions in the center mass system. This further broadens the distribution of the
directions of the recoil ions relative to the WIMP wind. Further,

(3) form factor effects decrease the momentum transfer whereas the correlations between the
direction of the recoiling ion and the original WIMP are maximal for maximal transfers.

All the above tend to randomize the recoil direction so that even if ideally it could be
exactly found experimentally it will serve as a weaker indicator for WIMPs.

For fully oriented crystalline detectors one can use  the twice-a-day modulations of the
WIMP signal due to the changing angle between the WIMP wind and the crystal planes with
the rotation of the earth.\cite{Avignone:2008cw}.

Four years earlier a similar idea had been suggested in Ref.~\cite{Sekiya}.  The advantage
of this suggestion is that the Stilbene crystal has low monoclinic symmetry and hence the
variation of the channeling efficiencies with time are likely to be large. This is not
the case for the highly symmetric cubic or diamond lattices of Na-I and Germanium where at
any time of day there are channeling through various 100, 110 and 111 crystalline planes
moderating the daily variations.  Still large scale such crystalline underground detectors
exist and searching these modulations in the data is certainly worthwhile.
The modulations occur almost twice a day. The reason is that most underground experiments
are within $\pm 3^o$ away from the $45^o$ latitude and the direction of the WIMP wind is
radially inward at $48^o$. Neglecting the $\sim 3^o$ discrepancies the $180^o$ degree
rotation of the earth occurring every twelve hours are, in fact,  12-hour rotations of
the earth and becomes a symmetry rotation around a 110 axis.

\section{Acknowledgements}

This minireview is based on lectures I gave at Princeton University and University of Maryland
College Park and inspired by a request of K.~K.~Phua.  The lectures presented work I did with
F.~T.~Avignone and R.~Creswick on the daily modulations and with I.~Yavin on the WIMP capture and
annihilation in the sun. I am very thankful to all of them and, in particular, to Itay~Yavin whose
familiarity with the new CDM models was of crucial help.


\begin{thebibliography}{0}
\bibitem{Begeman:1991iy}
K.~G.~Begeman, A.~H.~Broeils and R.~H.~Sanders,
Mon.\ Not.\ Roy.\ Astron.\ Soc.\  {\bf 249}, 523 (1991).

\bibitem{Koopmans:2002qh}
L.~V.~E.~Koopmans and T.~Treu,
Astrophys.\ J.\  {\bf 583}, 606 (2003)
[arXiv:astro-ph/0205281].

\bibitem{Markevitch:2003at}
M.~Markevitch {\it et al.},
Astrophys.\ J.\  {\bf 606}, 819 (2004)
[arXiv:astro-ph/0309303].

\bibitem{Komatsu:2008hk}
E.~Komatsu {\it et al.}  [WMAP Collaboration],
Astrophys.\ J.\ Suppl.\  {\bf 180}, 330 (2009)
[arXiv:0803.0547 [astro-ph]].

\bibitem{Chivukula:1992pn}
R.~S.~Chivukula, A.~G.~Cohen, M.~E.~Luke and M.~J.~Savage,
Phys.\ Lett.\  B {\bf 298}, 380 (1993)
[arXiv:hep-ph/9210274].

\bibitem{Nussinov:1985xr}
S.~Nussinov,
Phys.\ Lett.\  B {\bf 165}, 55 (1985).

\bibitem{Kaplan:2009ag}
D.~E.~Kaplan, M.~A.~Luty and K.~M.~Zurek,
arXiv:0901.4117 [hep-ph].

\bibitem{Foot:2007nn}
R.~Foot,
Int.\ J.\ Mod.\ Phys.\  A {\bf 22}, 4951 (2007)
[arXiv:0706.2694 [hep-ph]].

\bibitem{Witten:1984rs}
E.~Witten,
Phys.\ Rev.\  D {\bf 30}, 272 (1984).

\bibitem{Zhitnitsky:2006vt}
A.~Zhitnitsky,
Phys.\ Rev.\  D {\bf 74}, 043515 (2006)
[arXiv:astro-ph/0603064].

\bibitem{Forbes:2008zz}
M.~M.~Forbes and A.~R.~Zhitnitsky,
JCAP {\bf 0801}, 023 (2008).

\bibitem{Pospelov:2008jk}
M.~Pospelov, A.~Ritz and M.~B.~Voloshin,
Phys.\ Rev.\  D {\bf 78}, 115012 (2008)
[arXiv:0807.3279 [hep-ph]].

\bibitem{Kusenko:2008gh}
A.~Kusenko, B.~P.~Mandal and A.~Mukherjee,
Phys.\ Rev.\  D {\bf 77}, 123009 (2008)
[arXiv:0801.4734 [astro-ph]].

\bibitem{Boehm:2003bt}
C.~Boehm, D.~Hooper, J.~Silk, M.~Casse and J.~Paul,
Phys.\ Rev.\ Lett.\  {\bf 92}, 101301 (2004)
[arXiv:astro-ph/0309686].

\bibitem{Avignone:2009ay}
F.~T.~.~Avignone, R.~J.~Creswick and S.~Nussinov,
arXiv:0903.4451 [astro-ph.HE].

\bibitem{Finkbeiner:2009ug}
D.~P.~Finkbeiner, T.~Lin and N.~Weiner,
arXiv:0906.0002 [astro-ph.CO].

\bibitem{Drukier:1986tm}
A.~K.~Drukier, K.~Freese and D.~N.~Spergel,
Phys.\ Rev.\  D {\bf 33}, 3495 (1986).

\bibitem{Jungman:1995df}
G.~Jungman, M.~Kamionkowski and K.~Griest,
Phys.\ Rept.\  {\bf 267}, 195 (1996)
[arXiv:hep-ph/9506380].

\bibitem{Feng:2008ya}
J.~L.~Feng and J.~Kumar,
Phys.\ Rev.\ Lett.\  {\bf 101}, 231301 (2008)
[arXiv:0803.4196 [hep-ph]].

\bibitem{ArkaniHamed:2004fb}
N.~Arkani-Hamed and S.~Dimopoulos,
JHEP {\bf 0506}, 073 (2005)
[arXiv:hep-th/0405159].

\bibitem{Ji:2008cq}
X.~Ji, R.~N.~Mohapatra, S.~Nussinov and Y.~Zhang,
Phys.\ Rev.\  D {\bf 78}, 075032 (2008)
[arXiv:0808.1904 [hep-ph]].

\bibitem{Gondolo:2004sc}
P.~Gondolo, J.~Edsjo, P.~Ullio, L.~Bergstrom, M.~Schelke and E.~A.~Baltz,
JCAP {\bf 0407}, 008 (2004)
[arXiv:astro-ph/0406204].

\bibitem{TuckerSmith:2001hy}
D.~Tucker-Smith and N.~Weiner,
Phys.\ Rev.\  D {\bf 64}, 043502 (2001)
[arXiv:hep-ph/0101138].

\bibitem{Finkbeiner:2007kk}
D.~P.~Finkbeiner and N.~Weiner,
Phys.\ Rev.\  D {\bf 76}, 083519 (2007)
[arXiv:astro-ph/0702587].

\bibitem{Hisano:2004ds}
J.~Hisano, S.~Matsumoto, M.~M.~Nojiri and O.~Saito,
Phys.\ Rev.\  D {\bf 71}, 063528 (2005)
[arXiv:hep-ph/0412403].

\bibitem{Cirelli:2005uq}
M.~Cirelli, N.~Fornengo and A.~Strumia,
Nucl.\ Phys.\  B {\bf 753}, 178 (2006)
[arXiv:hep-ph/0512090].

\bibitem{ArkaniHamed:2008qn}
N.~Arkani-Hamed, D.~P.~Finkbeiner, T.~R.~Slatyer and N.~Weiner,
Phys.\ Rev.\  D {\bf 79}, 015014 (2009)
[arXiv:0810.0713 [hep-ph]].

\bibitem{Cholis:2008vb}
I.~Cholis, L.~Goodenough and N.~Weiner,
arXiv:0802.2922 [astro-ph].

\bibitem{Strassler:2006im}
M.~J.~Strassler and K.~M.~Zurek,
Phys.\ Lett.\  B {\bf 651}, 374 (2007)
[arXiv:hep-ph/0604261].

\bibitem{Baumgart:2009tn}
M.~Baumgart, C.~Cheung, J.~T.~Ruderman, L.~T.~Wang and I.~Yavin,
JHEP {\bf 0904}, 014 (2009)
[arXiv:0901.0283 [hep-ph]].

\bibitem{Cheung:2009qd}
C.~Cheung, J.~T.~Ruderman, L.~T.~Wang and I.~Yavin,
arXiv:0902.3246 [hep-ph].

\bibitem{Katz:2009qq}
A.~Katz and R.~Sundrum,
JHEP {\bf 0906}, 003 (2009)
[arXiv:0902.3271 [hep-ph]].

\bibitem{Adriani:2008zr}
O.~Adriani {\it et al.}  [PAMELA Collaboration],
Nature {\bf 458}, 607 (2009)
[arXiv:0810.4995 [astro-ph]].

\bibitem{Meade:2009rb}
P.~Meade, M.~Papucci and T.~Volansky,
arXiv:0901.2925 [hep-ph].

\bibitem{:2008zzr}
J.~Chang {\it et al.},
Nature {\bf 456}, 362 (2008).

\bibitem{Abdo:2009zk}
A.~A.~Abdo {\it et al.}  [The Fermi LAT Collaboration],
Phys.\ Rev.\ Lett.\  {\bf 102}, 181101 (2009)
[arXiv:0905.0025 [astro-ph.HE]].

\bibitem{Adriani:2008zq}
O.~Adriani {\it et al.},
Phys.\ Rev.\ Lett.\  {\bf 102}, 051101 (2009)
[arXiv:0810.4994 [astro-ph]].

\bibitem{Finkbeiner:2009mi}
D.~P.~Finkbeiner, T.~R.~Slatyer, N.~Weiner and I.~Yavin,
arXiv:0903.1037 [hep-ph].

\bibitem{Delaunay:2008pc}
C.~Delaunay, P.~J.~Fox and G.~Perez,
JHEP {\bf 0905}, 099 (2009)
[arXiv:0812.3331 [hep-ph]].

\bibitem{Nussinov:2009ft}
S.~Nussinov, L.~T.~Wang and I.~Yavin,
arXiv:0905.1333 [hep-ph].

\bibitem{Desai:2004pq}
S.~Desai {\it et al.}  [Super-Kamiokande Collaboration],
Phys.\ Rev.\  D {\bf 70}, 083523 (2004)
[Erratum-ibid.\  D {\bf 70}, 109901 (2004)]
[arXiv:hep-ex/0404025].

\bibitem{Lewin:1995rx}
J.~D.~Lewin and P.~F.~Smith,
Astropart.\ Phys.\  {\bf 6}, 87 (1996).

\bibitem{Ahmed:2008eu} Ahmed, 2008 eu

\bibitem{Ahmed:2009rh}
Z.~Ahmed {\it et al.}  [CDMS Collaboration],
arXiv:0907.1438 [astro-ph.GA].

\bibitem{Bottino:2007qg}
A.~Bottino, F.~Donato, N.~Fornengo and S.~Scopel,
Phys.\ Rev.\  D {\bf 77}, 015002 (2008)
[arXiv:0710.0553 [hep-ph]].

\bibitem{Chang:2008gd}
S.~Chang, G.~D.~Kribs, D.~Tucker-Smith and N.~Weiner,
Phys.\ Rev.\  D {\bf 79}, 043513 (2009)
[arXiv:0807.2250 [hep-ph]].

\bibitem{Ambrosio:1997tc}
M.~Ambrosio {\it et al.}  [MACRO Collaboration],
Astropart.\ Phys.\  {\bf 7}, 109 (1997).

\bibitem{Drobyshevski:2007zj}
E.~M.~Drobyshevski,
Mod.\ Phys.\ Lett.\  A {\bf 23}, 3077 (2008)
[arXiv:0706.3095 [physics.ins-det]].

\bibitem{Avignone:2008cw}
F.~T.~.~Avignone, R.~J.~Creswick and S.~Nussinov,
arXiv:0807.3758 [hep-ph].

\bibitem{Sekiya} H. Sekiya,
arXiv astro-ph/0405598
\end{thebibliography}
\end{document}